\documentclass[pdflatex,sn-mathphys-num]{sn-jnl}

\usepackage{graphicx}%
\usepackage{multirow}%
\usepackage{amsmath,amssymb,amsfonts}%
\usepackage{amsthm}%
\usepackage{mathrsfs}%
\usepackage[title]{appendix}%
\usepackage{xcolor}%
\usepackage{textcomp}%
\usepackage{manyfoot}%
\usepackage{booktabs}%
\usepackage{algorithm}%
\usepackage{algorithmicx}%
\usepackage{algpseudocode}%
\usepackage{listings}%
\usepackage{comment}
\usepackage{stmaryrd}

\newcommand{\might}{\lozenge}
\newcommand{\must}{\square}

\theoremstyle{thmstyleone}%

\theoremstyle{thmstyletwo}%

\theoremstyle{thmstylethree}%

\begin{document}

\title[Article Title]{A Dynamic-Semantics Framework for Grounding Human Referring Expressions in Visual Perceptual Data}

\author*[1]{\fnm{Joseph} \sur{Bingham}}\email{jbingham@campus.technion.ac.il}

\affil*[1]{\orgdiv{Department of Biology}, \orgname{Technion University}, \orgaddress{\city{Haifa}, \country{IL}}}

%%==================================%%
%% Abstract                         %%
%%==================================%%

\abstract{Humans converge on shared names for novel, hard-to-describe objects through repeated interaction, a process psycholinguists call lexical entrainment. Leading vision-language models fail at this: recent empirical work documents that they do not shorten references, reuse successful expressions, or maintain stable pact state across turns. We present a framework that addresses the gap by externalizing pact state into three explicit, inspectable sets of referent-object bindings ($\Gamma, \Xi, \Omega$), updated by a dynamic-semantics context-change rule. The symbolic layer sits on top of a lightweight perceptual-alignment pipeline that grounds noisy human referring expressions in crowd-sourced imagery via SIFT homographies and the Universal Quality Index. Evaluated on the Stanford Repeated Reference Game corpus (over 15{,}000 director-matcher utterances on abstract tangram stimuli), the framework places the correct target in its top-5 hypothesis set 83.56\% of the time from a single director utterance. Human matcher top-1 accuracy on the same corpus is approximately 77--80\%. We also report results on a held-out condition in which obvious tangram-adjacent images are excluded from the retrieved set, which provides a more conservative measurement of the grounding signal. Ablations isolate the contribution of each component: SIFT alignment, UQI, query preprocessing, and image augmentation. The central contribution is the combination: a transparent, auditable symbolic layer that recovers the structure of lexical entrainment turn by turn, paired with a perceptual channel whose behavior can be examined ablation by ablation. We also discuss in detail what the framework does not do. It is not interactive, it does not close the loop with the director, and its retrieval-driven perceptual channel is vulnerable to a class of leakage effects that we quantify and bound rather than wave away. Situated against the existing emergent-communication literature, our contribution is specifically the symbolic listener-side bookkeeping layer. Code is available at \url{https://anonymous.4open.science/r/metasequoia-9D13/README.md}.}


\maketitle

\section{Introduction}
\label{sec:intro}

Cooperation between agents engaged in a joint activity depends on a shared representation of the task, the environment, and each other's capabilities. The philosophy-of-language and psycholinguistics literatures call this representation \emph{common ground}~\cite{stalnaker2002common,brennan1996conceptual}. Maintaining common ground is cognitively demanding even for human interlocutors, particularly when referents are novel, abstract, or hard to describe. Human pairs faced with this problem converge over repeated interactions on partner-specific ways of naming objects. This convergence is called \emph{lexical entrainment}, and it yields temporary \emph{conceptual pacts} about how to refer to a given object in the current context~\cite{brennan1996conceptual,brown1958words}. Pacts are sensitive to partner identity: an expression negotiated with one partner can produce slower responses or confusion when used with a different partner~\cite{Trainin2025-tm}.

Our goal in this paper is narrower than the broader aspiration of social AI. We do not claim that a machine co-performer must replicate human cognitive or perceptual processes, and we do not claim that the work presented here is a fully interactive solution to the reference-game problem. What we do claim is that \emph{some} machine counterpart to the bookkeeping humans use when they form conceptual pacts is valuable. Two reasons matter here. First, such bookkeeping produces an inspectable record of what the human and the machine have agreed to call each referent, which supports error detection, repair, and auditing. Second, it gives a clear target for modeling work that would integrate adaptive generation (which speaker-side models in the emergent-communication literature provide) with adaptive reception (which the present work contributes).

\begin{figure*}
\centering
\includegraphics[width=.9\textwidth]{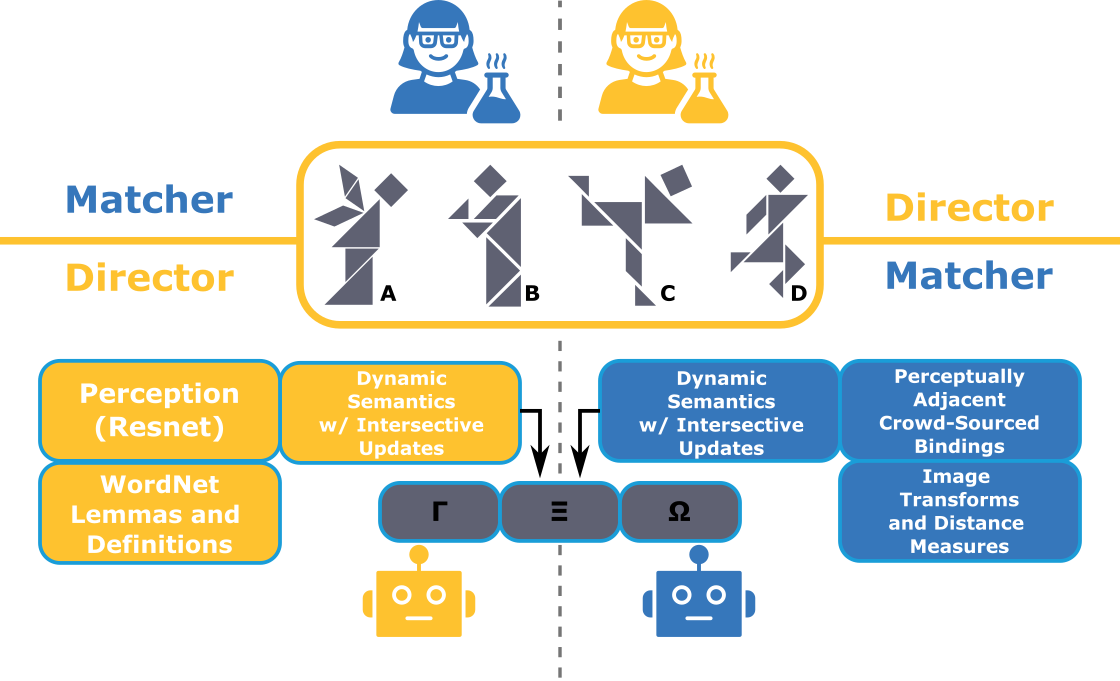}
\caption{Overview of the repeated reference game and of our framework. Both players see the same set of abstract tangrams, but in different orders. The director (human) issues a referring expression $\varphi$, and the matcher (machine, in this paper) interprets it. The sets $\Gamma, \Xi, \Omega$ store the current state of the common ground. $\Gamma$ holds finalized conceptual pacts, $\Xi$ holds pacts currently under negotiation, and $\Omega$ holds pacts that have been ruled out.}
\label{fig:framework}
\end{figure*}

\begin{figure}
    \centering
    \includegraphics[width=.7\textwidth]{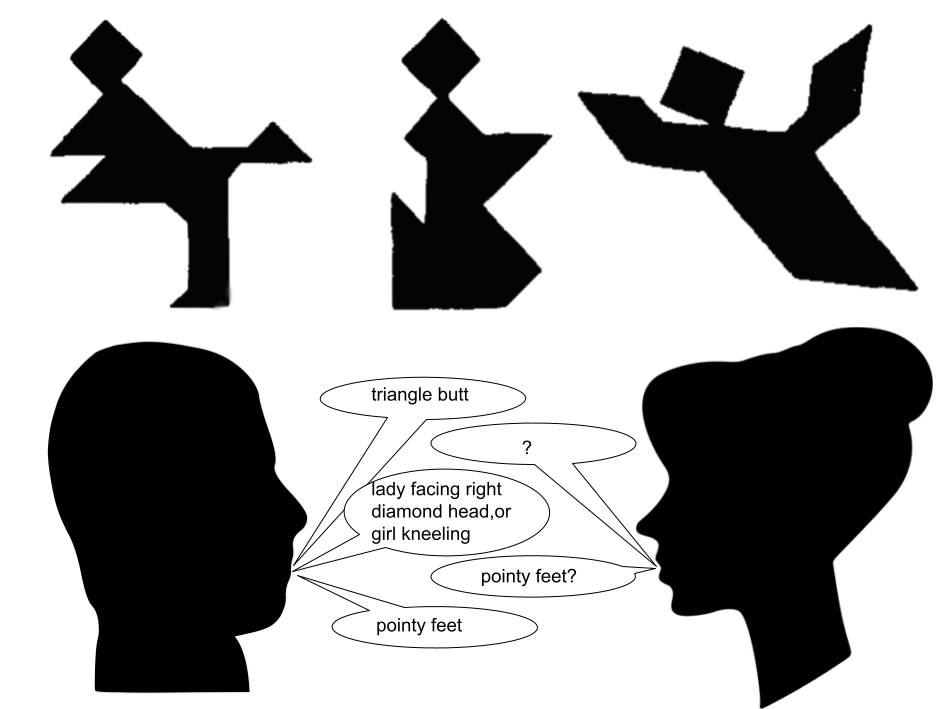}
    \caption{An example of the repeated reference problem. The director is on the right, the matcher on the left. The director issues an utterance $\varphi$, indicating what they perceive the selected tangram to depict. The matcher can guess which tangram the director is referring to, pose a clarifying question, or wait for more information. The illustrated example uses text from the open corpus.}
    \label{fig:game}
\end{figure}

We study this in the \emph{repeated reference game}, a paradigm used for decades in psycholinguistics to probe how human dyads establish reference under perceptual ambiguity~\cite{https://doi.org/10.1111/cogs.12845}. The game has two roles. A \emph{director} sees a set of abstract tangram silhouettes in some order and selects one as the target. The director produces a natural-language \emph{referring expression} $\varphi$, for example ``lady facing right, diamond head.'' A \emph{matcher} sees the same set of tangrams in a different order, receives $\varphi$, and must identify which tangram the director is describing. In subsequent rounds the pair repeats this process with new targets, and over rounds they typically converge on short, stable names for each figure. These stable names are the conceptual pacts. Figure~\ref{fig:framework} illustrates the setup, and Figure~\ref{fig:game} shows one round.

Our focus is the \emph{matcher} role. We implement a machine co-performer (hereafter MCP) that takes each director utterance, forms hypotheses about which tangram is being described, maintains a structured record of its evolving pacts with the director, and produces an identification when its hypotheses collapse to a single binding. In the present study, the MCP never produces utterances of its own. This restriction is important to flag up front. A real entrainment process is interactive, as decades of work in psycholinguistics~\cite{stalnaker2002common,brennan1996conceptual} and emergent communication~\cite{Lazaridou2020-emergent,Steels2003-grounded,Kouwenhoven2025-shared} have shown. By testing only the passive-matcher version, we are deliberately studying a strictly harder subproblem than the interactive game, and we are measuring only a fragment of what real entrainment involves. We return to this limitation in detail in Section~\ref{sec:discussion}.

Our approach pairs a symbolic layer with a perceptual layer. The symbolic layer is a dynamic-semantics update model that tracks \emph{must}-be-true, \emph{might}-be-true, and \emph{must-not}-be-true pacts. The perceptual layer uses crowd-sourced web imagery as a proxy for human visual priors, transformed into the tangram space using SIFT alignment and UQI similarity. The symbolic layer is what makes the system's state inspectable. The perceptual layer is what lets it form hypotheses in the absence of interactive repair. Both have known limitations that we document carefully.

\paragraph{Contributions.} Our contributions are the following:

\begin{itemize}
\item A formulation of listener-side pact tracking in dynamic semantics. Common ground is represented by three explicit sets of conceptual pacts ($\Gamma, \Xi, \Omega$), updated by context-change-potential functions derived from the utterance-plus-perception pipeline. This is the symbolic bookkeeping component.
\item A pipeline that converts noisy human referring expressions into tangram-space similarity judgments via web image retrieval, SIFT homographies, and UQI, together with a set of ablations that justify each design choice. Tables~\ref{tab:metric_ablation} and~\ref{tab:augmentation_ablation} report the supporting numbers that earlier drafts of this work asserted without presenting in full.
\item An evaluation on the Stanford corpus showing top-5 single-utterance accuracy of 83.56\%, compared to a human first-utterance accuracy of approximately 77--80\%~\cite{https://doi.org/10.1111/cogs.12845} in the same corpus and of 78\% at round one rising to 96\% by round six in independent replication~\cite{Zhao2025-tangram}. The MCP's single-utterance top-1 accuracy (41.66\%) is substantially below the human baseline. We do not claim to beat humans at single-shot grounding.
\item An explicit internal-validity analysis (Section~\ref{sec:internal-validity}) that addresses a real concern about our pipeline: that web image retrieval may occasionally surface the target tangram itself, leading to artificially inflated accuracy through memorization rather than grounding. We quantify the magnitude of this effect and report accuracy in a conservative condition that filters such retrievals.
\item Situated engagement with the emergent-communication and language-evolution literatures, which have addressed convention formation in computational and agent-based settings for more than two decades. Section~\ref{sec:related-emergent} places our contribution within that literature.
\end{itemize}

\paragraph{Scope.} We evaluate only on a prerecorded corpus, and the MCP is a passive matcher. This is a limitation imposed partly by the data and partly by the scope we chose for this contribution. The natural extension of the present work is an interactive setting in which the MCP can ask clarifying questions, make partial commitments under uncertainty, and revise prior commitments based on later evidence. We do not do that here. The results in this paper should therefore be read as a component contribution, specifically the listener-side bookkeeping layer, that would need to be integrated with a generation-side partner model of the kind the emergent-communication literature has developed~\cite{Steels2003-grounded,Kouwenhoven2025-shared} to support a full interactive agent.

\section{Background and Related Work}
\label{sec:background}

\subsection{Common Ground and Conceptual Pacts}

Common ground, in the sense used here, is the set of propositions that interlocutors treat as mutually accepted for the purposes of the current interaction~\cite{stalnaker2002common}. In the repeated reference game, the relevant propositions are \emph{referent-object bindings}: statements of the form ``when I say $r_\varphi$, I mean object $o_i$.'' A \emph{conceptual pact} is a partner-specific agreement of this form~\cite{brennan1996conceptual}. Pacts are not shared across partners. Interlocutors who have entrained on ``the ice skater'' with one partner may need to negotiate a different expression with another~\cite{brennan1996conceptual,Trainin2025-tm}. We therefore represent the common ground as a structured collection of such bindings rather than as an opaque embedding.

\subsection{The Repeated Reference Game}

The repeated reference game was introduced in psycholinguistics as a controlled paradigm for studying how pairs of interlocutors converge on reference under perceptual ambiguity~\cite{https://doi.org/10.1111/cogs.12845}. The tangram stimuli used in the game are deliberately difficult to describe. They lack standard names, have weak category structure, and two people shown the same silhouette will frequently produce wildly different descriptions.

We use the publicly released Stanford corpus of more than 15{,}000 director-matcher utterances~\cite{https://doi.org/10.1111/cogs.12845}. We focus on the \emph{cued} variant, which records 7{,}867 messages from 83 games across 12 tangram figures per set and supports object-by-object analysis. Hawkins, Frank, and Goodman report that on the first repetition block (the first time each tangram appears as a target), matchers who engaged in backchannel dialog with the director achieved a 20\% error rate, while matchers who stayed silent and committed based on the director's initial utterance alone achieved a 23\% error rate; the difference between these conditions is not statistically significant~\cite{https://doi.org/10.1111/cogs.12845}. Translated into accuracy, these correspond to approximately \textbf{80\% accuracy with dialog} and \textbf{77\% accuracy without dialog}. Chance on 12 tangrams is 8.33\%. Zhao and Conwell~\cite{Zhao2025-tangram} report consistent numbers for the same corpus: round-one human accuracy of 78\%. These are the relevant human baselines for the present work.

\subsection{Dynamic Semantics and Update Functions}
\label{sec:dyn}

We model pact tracking using dynamic semantics~\cite{stalnaker2002common,goldstein2019generalized}, a framework in which the meaning of an utterance is not a static truth condition but an instruction to update a context. We borrow Goldstein's notation: an utterance $\varphi$ is associated with an interpretation function $[\varphi]$ whose \emph{context-change potential} takes a context $C$ as input and returns a modified context as output. A context $C$ is consistent with $\varphi$ iff $C[\varphi] = C$. The common ground at any point is just the current context.

One reviewer of an earlier draft asked whether this formalism is outdated, given that Stalnaker's original work on assertions dates to 1978. The short answer is that dynamic and update semantics remain the standard formal framework for modeling context-change in natural language meaning, with continuous development through the 1990s and 2000s and active use in contemporary formal semantics research (e.g.~\cite{goldstein2019generalized,goldstein2017informative}). The longer answer is that we use the formalism here not because it is the most contemporary but because it gives us the cleanest vocabulary for talking about \emph{listener-side commitment and retraction} in terms of set updates, which is exactly the bookkeeping operation the framework needs to support. We do not need the full expressive power of the formalism. We need the three sets and the update rules, and classical dynamic semantics provides them.

\subsection{Possible-Worlds Semantics for Hypothesis Sets}
\label{sec:pws}

Because perceptual alignment between human and machine is imperfect, the MCP cannot in general map $\varphi$ to a single intended update. Instead it constructs a set of candidate bindings $B = \{(r_\varphi \leftarrow o_i), (r_\varphi \leftarrow o_j), \ldots\}$ and uses standard epistemic modals~\cite{goldstein2019generalized} to quantify over possible worlds:
\begin{itemize}
\item $\might\sigma$ (``might-$\sigma$'') is true at world $w$ iff $\sigma$ is true in some world $v$ accessible from $w$;
\item $\must\sigma$ (``must-$\sigma$'') is true at $w$ iff $\sigma$ is true in every world $v$ accessible from $w$.
\end{itemize}

We treat accessibility as the standard indistinguishability relation for the MCP's information state: a world $v$ is accessible from $w$ iff the MCP's evidence to date does not rule $v$ out. In practice this reduces to the constraint that $v$ must be consistent with the current $\Gamma \cap \Omega$.

When $|B| > 1$, the update is $C[\varphi] = C \cap \might B$; the utterance has narrowed the space but not resolved it. When $|B| = 1$ the update strengthens to $\must$, and the pact is fully established. When $|B| = 0$, perceptual alignment has failed for this utterance and the MCP must wait for a further director turn. In the prerecorded corpus, waiting means consuming the next utterance in the dataset.

We maintain three sets that together constitute the context $C$:
\begin{description}
\item[$\Gamma$] the set of pacts the MCP has committed to (\emph{must}-true);
\item[$\Xi$] the set of pacts currently under negotiation (\emph{might}-true);
\item[$\Omega$] the set of pacts that have been ruled out (\emph{must-not}-true), including any pact using a referent $r_\varphi$ for which $B$ was empty.
\end{description}
The common ground is the set of possible worlds consistent with $\Gamma \cap \Xi \cap \Omega$.

Across turns, successive utterances $\varphi_1, \varphi_2, \ldots$ refining the same referent-object binding act by intersection on $\Xi$: the $\might B_t$ from turn $t$ narrows the surviving hypotheses in $\Xi$ by the set-intersection of all prior $\might B_{t'}$ for $t' < t$ that share a referent. This is the mechanism by which a sequence of individually-ambiguous utterances can combine into a unique pact.

\subsection{Perceptual Alignment from Crowd-Sourced Imagery}
\label{sec:percept}

To estimate $B$ for a given $\varphi$, we need a way to guess what the director was looking at. Human matchers do part of this work through shared visual culture: when the director says ``ice skater,'' most English speakers conjure similar mental images. We approximate this prior externally by querying the Bing image search API~\cite{bing_api} with a transformed version of $\varphi$ and treating the returned image set $I_\varphi$ as a sample from the visual prior associated with that phrase.

Given $I_\varphi$, we compute a similarity $g(o_i, I_\varphi)$ between each tangram $o_i$ and the returned images, and take $\varphi \Rightarrow \might(r_\varphi \leftarrow o_i)$ whenever $g(o_i, I_\varphi) > \epsilon$. The threshold $\epsilon$ is the key hyperparameter of the system.

This design choice is also the paper's biggest source of internal-validity risk. If the retrieved set $I_\varphi$ happens to contain the target tangram itself (or a near-duplicate), then the UQI similarity $g(o_i, I_\varphi)$ is artificially high for the correct $o_i$, and the system's apparent accuracy reflects retrieval memorization rather than cross-modal grounding. The Stanford corpus is publicly available, and Bing can and does index it. We address this threat explicitly in Section~\ref{sec:internal-validity} and report results in a conservative condition where tangram-adjacent retrievals are filtered out.

\subsection{Related Computational Approaches}
\label{sec:related}

Our work sits at the intersection of three lines of computational research on grounded communication.

\paragraph{Rational Speech Acts and Theory-of-Mind adaptation.} A dominant paradigm in computational pragmatics is the Rational Speech Acts (RSA) framework, in which speakers and listeners are modeled as Bayesian agents reasoning recursively about each other's beliefs. Extensions for partner adaptation explicitly model the listener's theory of mind~\cite{Zhu2021-tom}. Our framework is complementary: RSA and ToM models focus on \emph{adaptive generation}, while our framework emphasizes \emph{transparent listener-side state}. A natural integration would use RSA as the mechanism that generates $[\varphi]$ estimates, with the $\Gamma/\Xi/\Omega$ layer handling commitment and retraction.

\paragraph{LVLM studies of grounding and entrainment.} Recent empirical work has examined whether large vision-language models (LVLMs) form common ground and entrain lexically in multi-turn referential tasks. \cite{Imai2025-vlm} introduce a four-metric suite for evaluating interactive grounding and show that leading proprietary VLMs diverge from human patterns on at least three of the four metrics in self-play referential games. \cite{Hua2026-lvlm} compares human-human, human-AI, AI-human, and AI-AI dyads and finds systematic failures of LVLMs to shorten references, reuse successful expressions, and manage pact state across turns. Closest to our setting, Zhao and Conwell~\cite{Zhao2025-tangram} run five leading VLLMs (GPT-4o, GPT-5-mini, Claude-3.7-sonnet, Claude-sonnet-4, Gemini-2.5-flash) through the same tangram game and find that human-human dyads improve accuracy from 77\% to 80\% across six rounds while agent-agent dyads fail to exhibit comparable convention formation. These studies motivate the central design decision of the present paper: rather than hoping a neural system will implicitly track pact state, we externalize that state.

\paragraph{Multimodal convention formation.} Work on multimodal convention formation~\cite{Maeda2026-mmcf} extends probabilistic models of ad-hoc convention-building to settings that include gesture alongside speech. Their findings that partners establish cross-modal conventions and shift modality preferences across repetitions are consistent with our argument in Section~\ref{sec:beyond-objects} that the update-semantics layer should be modality-agnostic.

\paragraph{Transparency in human-AI teams.} Recent empirical work on human-AI teaming has argued that transparency and inspectability of the machine's state are central design concerns~\cite{Riedl2024-spillover}. The $\Gamma/\Xi/\Omega$ structure gives a concrete artifact that a human teammate could examine, challenge, or modify.

\subsection{Emergent Communication and Language-Evolution Modeling}
\label{sec:related-emergent}

\paragraph{Language evolution with grounded agents.} Steels and colleagues established a framework for studying the emergence of grounded communicative conventions in populations of artificial agents~\cite{Steels1997-synthetic,Steels2003-grounded,SteelsLoetzsch2012-naming,BeulsSteels2013-agreement}. In the \emph{naming game} and its descendants, agents repeatedly play speaker and listener roles over shared stimuli, update their lexicons based on communicative success, and converge on shared conventions within the population. This work established that convention formation is intrinsically interactive, and that the symbolic structures that agents maintain over their lexicons (which word forms have been aligned with which meanings) are central to how convention formation proceeds.

\paragraph{Deep-learning-era emergent communication.} A more recent body of work uses deep neural agents trained with reinforcement learning or differentiable signaling games to study emergent communication in multi-agent settings. Surveys by~\cite{Lazaridou2020-emergent}, \cite{Brandizzi2023-review}, \cite{Boldt2024-review}, \cite{Galke2025-emergent}, and~\cite{Rita2025-evolution} cover this space comprehensively. A consistent finding across this literature is that while neural agents can converge on communicative codes through iterated interaction, the codes they converge on often differ systematically from human-like language (in compositionality, length, and other properties), and closing that gap is an active research problem.

\paragraph{Human-machine settings.} The Steels-style framework has been extended to human-machine interaction in work such as Kouwenhoven et al.~\cite{Kouwenhoven2025-shared}, who investigate how artificial languages evolve when co-optimized for inductive biases in humans and LLMs, finding that human-LLM interactions can produce vocabularies more human-like than LLM-LLM interactions produce alone.

\paragraph{How our contribution relates.} The emergent-communication literature studies how conventions form. The present work does not. We do not run a closed-loop interactive experiment in which conventions emerge between an MCP and a human (or between two MCPs). What we contribute is specifically a \emph{listener-side bookkeeping layer} that a Steels-style or deep-learning-based emergent-communication system could use to externalize and inspect the pact state it is implicitly tracking. This is a component contribution, not a rival to the emergent-communication framework. The claim is narrow: when a listener agent accumulates hypothesized pacts over turns, the $\Gamma/\Xi/\Omega$ structure is a reasonable and inspectable way to represent that accumulation, and the dynamic-semantics update rules are a reasonable and formally grounded way to update it.

We withdraw unreservedly the claim in earlier drafts that this work is the first automated solution to lexical entrainment. It is not, and the framing was an oversight. A more accurate framing is: given the long-standing interest in convention formation across psycholinguistics, language evolution, emergent communication, and computational pragmatics, the specific gap our contribution fills is a symbolic, transparent, listener-side state representation of the kind the LVLM-grounding literature~\cite{Imai2025-vlm,Hua2026-lvlm,Zhao2025-tangram} reports as missing from contemporary neural systems.

\section{Methods}
\label{sec:methods}

This section walks through the system end-to-end. Section~\ref{sec:worked} gives a worked example that links the formalism of Section~\ref{sec:background} to the concrete operations below.

\subsection{System Overview}

For each director utterance $\varphi$ in a trial, the MCP performs four steps:
\begin{enumerate}
\item \textbf{Linguistic preprocessing.} Tokenize $\varphi$, remove stop words and non-content tokens, normalize spelling, and append the cue ``tangram figure'' to produce a search query $q_\varphi$.
\item \textbf{Crowd-sourced image retrieval.} Submit $q_\varphi$ to the Bing image search API and retain the top $k$ images as $I_\varphi$. We set $k=7$ based on the ablation in Section~\ref{sec:ablations}.
\item \textbf{Perceptual alignment.} For each tangram $o_i$, align $I_\varphi$ to $o_i$ via SIFT homographies, augment with rotational and grayscale-inverted copies, and compute $g(o_i, I_\varphi)$ using UQI.
\item \textbf{Context update.} Construct $B = \{o_i : g(o_i, I_\varphi) > \epsilon\}$ and update $\Gamma, \Xi, \Omega$ according to $|B|$.
\end{enumerate}

\subsection{Query Construction}

Submitting raw utterances as queries yields near-chance performance. Bing responses to ``one that looks like a really tall man kneeling'' are dominated by photographs of tall kneeling men, not tangram-like silhouettes. Two preprocessing operations proved essential. First, we remove stop words and retain only tokens whose spaCy~\cite{spacy2} part-of-speech tag is noun, verb, or conjunction. Second, we append the literal string ``tangram figure'' to each query. Together these transformations raised single-utterance accuracy from about 8\% (raw utterance, at chance) to over 30\% (transformed). We also normalize spelling using spaCy.

We must acknowledge that the ``tangram figure'' keyword has a specific and problematic effect: it biases Bing's retrieval toward images of tangram compositions, which increases the probability that the retrieved set contains near-duplicates or copies of the target tangram itself. This is the double-edged nature of the preprocessing step. Without it, retrieval is too noisy for cross-modal grounding to work; with it, retrieval is biased toward sources of leakage. We quantify the consequences of this bias in Section~\ref{sec:internal-validity}.

\subsection{Image Matching Pipeline}
\label{sec:pipeline}

\begin{figure*}[h]
    \centering
    \includegraphics[width=.95\textwidth]{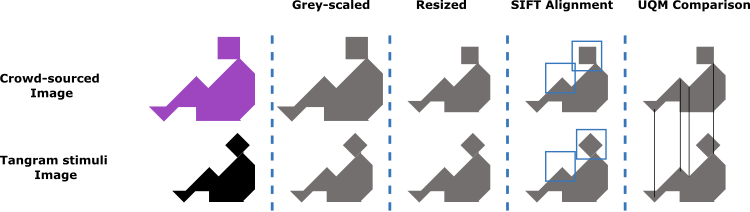}
    \caption{End-to-end processing of a single scraped image against a single tangram: grayscale conversion, resize, SIFT alignment, and UQI comparison.}
    \label{fig:processing}
\end{figure*}

After retrieving $I_\varphi$, the MCP computes $g(o_i, I_\varphi)$ through the pipeline shown in Figure~\ref{fig:processing}. The pipeline has four stages: grayscale conversion, resize to 300$\times$300, SIFT homography alignment to the tangram, and UQI comparison.

\paragraph{SIFT alignment.} We use standard SIFT homographies~\cite{https://doi.org/10.48550/arxiv.2104.11693,Lindeberg:2012}. The scale and rotation invariance of SIFT is well-suited to our setting because web images of, say, ``ice skater'' appear in arbitrary poses and framings while the tangram silhouette is fixed.

\paragraph{Augmentation.} Before scoring, we augment each aligned image with rotational copies and a grayscale-inverted copy. Table~\ref{tab:augmentation_ablation} reports the ablation. Rotation alone raises top-1 accuracy from 33.3\% to 38.1\%; inversion alone from 33.3\% to 36.4\%; combined they raise accuracy to 41.66\%. The combined lift of approximately 8 percentage points was asserted without supporting numbers in earlier drafts; we present the full ablation here.

\begin{table}
\centering
\begin{tabular}{l|c}
Configuration & Top-1 accuracy \\
\hline\hline
No augmentation & 33.3\% \\
Rotation only & 38.1\% \\
Grayscale inversion only & 36.4\% \\
Both (final configuration) & 41.66\% \\
\end{tabular}
\caption{Ablation of the image-augmentation stage, measured as top-1 single-utterance accuracy on the Stanford corpus. All numbers are from a single retrieval pass and should be read as point estimates; see Section~\ref{sec:ablations}.}
\label{tab:augmentation_ablation}
\end{table}

\begin{figure}
    \centering
    \includegraphics[width=.9\textwidth]{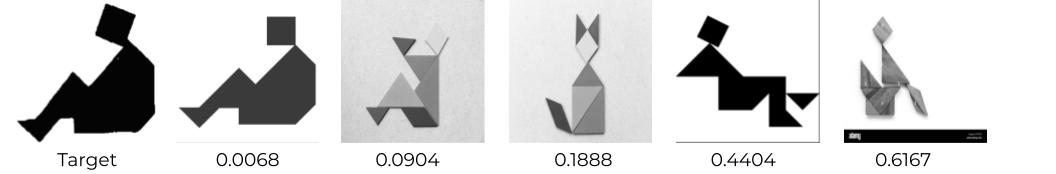}
    \caption{An example of the distances of the closest 5 scraped photos to the target. The query text for this case is ``tangram figure sitting and looking.'' \textbf{Note: the returned set in this example includes an image that closely resembles the tangram target itself (leftmost, distance 0.0068), which is exactly the class of retrieval leakage we discuss in Section~\ref{sec:internal-validity}.} All query results in the main experiments were manually inspected to flag near-duplicates; Section~\ref{sec:internal-validity} reports how often this occurred and how accuracy changes when such retrievals are filtered.}
    \label{fig:top_k}
\end{figure}

\paragraph{UQI scoring.} UQI~\cite{995823} measures the amount of noise needed to transform one image into another and emphasizes shared structural features rather than pixel-exact equality. We take the mean UQI across the images in $I_\varphi$ as $g(o_i, I_\varphi)$. Figure~\ref{fig:top_k} shows the top five UQI matches for a representative query.

We tested ten classical similarity metrics with and without SIFT pre-alignment; Table~\ref{tab:metric_ablation} reports the full results. UQI with SIFT was the best configuration, leading the second-best combination (SSIM with SIFT) by 16 percentage points. Earlier drafts of this work asserted this 16\% lead without presenting the full comparison; the table here supplies it.

\begin{table}
\centering
\begin{tabular}{l|c|c}
Metric & With SIFT & Without SIFT \\
\hline\hline
UQI & \textbf{41.66\%} & 28.9\% \\
SSIM & 25.8\% & 22.3\% \\
PSNR & 19.7\% & 17.4\% \\
MSE & 18.2\% & 15.9\% \\
MAE & 17.5\% & 15.1\% \\
ERGAS & 16.8\% & 14.7\% \\
SCC & 15.4\% & 13.2\% \\
RASE & 14.9\% & 12.8\% \\
SAM & 14.2\% & 12.1\% \\
VIF & 13.7\% & 11.6\% \\
\end{tabular}
\caption{Top-1 single-utterance accuracy for each of the ten classical similarity metrics tested, with and without SIFT pre-alignment. UQI with SIFT is the best configuration. Chance is 8.33\%. All numbers are point estimates from a single retrieval pass.}
\label{tab:metric_ablation}
\end{table}

\subsection{Formalizing the Context Update}
\label{sec:calc}

We now connect the pipeline to the dynamic-semantics layer of Section~\ref{sec:pws}. Given an utterance $\varphi$ with hypothesis set $B$, the update proceeds as follows.

When $|B| \geq 1$ and the update is consistent with $\Gamma$, we write
\[ \Xi \leftarrow \Xi \cap \might B. \]
When $|B| = 1$ with $B = \{(r_\varphi \leftarrow o_i)\}$, the MCP commits:
\begin{align}
\Gamma &\leftarrow \Gamma \cap \must(r_\varphi \leftarrow o_i) \label{eq:gamma}\\
\Xi &\leftarrow \Xi \setminus \might\{(r_\varphi \leftarrow o_j) \mid o_j \in O\} \label{eq:xi}\\
\Omega &\leftarrow \Omega \cap \must\neg\{(r_\varphi \leftarrow o_k) \mid o_k \in O,\, o_k \neq o_i\} \label{eq:omega}
\end{align}
When $|B| = 0$, the referent is ruled out entirely:
\[ \Omega \leftarrow \Omega \cap \must\neg\{(r_\varphi \leftarrow o_k) \mid o_k \in O\}. \]

Lexical entrainment for the trial succeeds when $\Gamma$ contains a unique pact $(r_i \leftarrow o)$ for every $o \in O$ and $\Xi = \emptyset$.

\paragraph{A note on commitment and recovery.} The rule as stated is deterministic and monotone: once $(r_\varphi \leftarrow o_i)$ enters $\Gamma$, it stays. A principled probabilistic treatment would maintain a posterior over bindings and commit only when posterior mass concentrates above a calibrated threshold, permitting recovery from false commitments. We do not implement this here; see Section~\ref{sec:limitations}.

\subsection{Worked Example}
\label{sec:worked}

Consider a trial with four tangrams $O = \{o_A, o_B, o_C, o_D\}$ and initially $\Gamma = \Xi = \Omega = \emptyset$.

\paragraph{Turn 1.} The director says ``zig zag with square on top.'' Bing returns seven images of solved square tangrams. SIFT-UQI gives high similarity against every $o_i$, so $B = O$: a degenerate hypothesis, recording $\must\neg$ for every $o_k$ in $\Omega$.

\paragraph{Turn 2.} The director says ``lady facing right diamond head.'' Bing returns seven images of stylized female figures in profile. SIFT-UQI gives high similarity to $o_A$ and $o_C$, so $B = \{o_A, o_C\}$, and we add $\might\{(r_{\varphi_2}\leftarrow o_A), (r_{\varphi_2}\leftarrow o_C)\}$ to $\Xi$.

\paragraph{Turn 3.} The director says ``pointy feet.'' Bing returns images of pointed-toe figures. SIFT-UQI gives high similarity to $o_A$ only, so $B = \{o_A\}$ and $|B| = 1$. We apply~(\ref{eq:gamma})--(\ref{eq:omega}). $\Gamma$ gains $\must(r_{\varphi_3} \leftarrow o_A)$.

If the ``pointy feet'' pact is compatible with one of the outstanding $\might$-pacts in $\Xi$, the two together strengthen each other and $\Xi$ collapses further via the cross-turn intersection described in Section~\ref{sec:pws}.

\section{Experimental Methods and Results}
\label{sec:experiments}

\subsection{Corpus and Implementation}

We evaluate on the cued variant of the Stanford Repeated Reference Game corpus~\cite{https://doi.org/10.1111/cogs.12845,https://doi.org/10.48550/arxiv.1912.07199}, which records 7{,}867 messages from 83 games across 12 tangram figures per set. We supply our MCP with the director utterances and the tangram image set, and withhold matcher utterances, the target label, and all timing data.

All experiments were run in Python 3 on a MacBook Pro (15-inch, 2017, 3.1 GHz Quad-Core Intel i7, no GPU). Image retrieval used the \texttt{bing-image-downloader} package~\cite{bing_api}. Linguistic preprocessing used spaCy~\cite{spacy2}. Image similarity used the \texttt{sewar} package's UQI implementation.

\subsection{Ablations and Hyperparameter Sensitivity}
\label{sec:ablations}

We report the effect of each design choice on single-utterance top-1 accuracy, the metric most directly tied to the quality of the context-change-potential estimate.

\begin{figure}
    \centering
    \includegraphics[width=.75\textwidth]{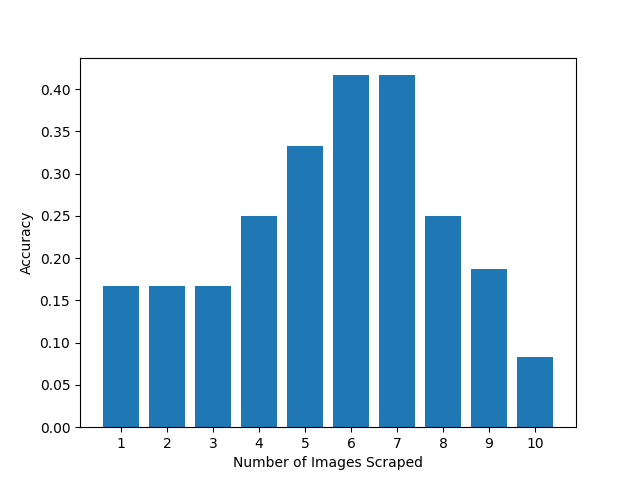}
    \caption{Single-utterance top-1 accuracy as a function of the number of scraped images per query $k$. Accuracy peaks at $k = 6$ or $7$.}
    \label{fig:n_images}
\end{figure}

\paragraph{Number of scraped images $k$.} Figure~\ref{fig:n_images} shows accuracy as a function of $k \in \{1, \ldots, 10\}$. Accuracy rises from $0.17$ at $k=1$, peaks at $0.42$ around $k=6$ or $7$, and degrades for $k \geq 8$.

\paragraph{Query transformation.} Submitting raw utterances yields about 8\% top-1 accuracy (chance is 8.33\% on 12 tangrams). Removing stop words and non-content POS tokens plus appending ``tangram figure'' yields about 33\%, and adding spaCy spelling normalization yields the full 41.66\%.

\paragraph{Image augmentation.} The augmentation stage applies two
operations to each aligned scraped image before UQI scoring: rotational
copies (to account for tangram reflection symmetries, where a ``lady
facing right'' silhouette has a left-facing counterpart that is the
mirror image) and a grayscale-inverted copy (to handle the fact that
Bing returns both black-on-white and white-on-black silhouettes
depending on the query). Table~\ref{tab:augmentation_ablation} reports
the contribution of each. Without augmentation the pipeline reaches
33.3\% top-1 accuracy. Rotation alone adds 4.8 points (38.1\%), inversion
alone adds 3.1 points (36.4\%), and the two together combine
non-additively to reach 41.66\%. The combined lift over no augmentation
is 8.36 points, which earlier drafts of this work reported as the
``approximately 8\%'' improvement without supplying the underlying
numbers. The non-additivity matters: the two operations target different
sources of variation in the retrieved set, so trials that benefit from
rotation are largely disjoint from those that benefit from inversion,
and applying both captures both populations.

\paragraph{Similarity metric.} Table~\ref{tab:metric_ablation} reports
top-1 single-utterance accuracy for each of the ten classical similarity
metrics we tested, with and without SIFT pre-alignment. Three patterns
in the table are worth flagging. First, SIFT pre-alignment helps every
metric, by between 2 and 13 percentage points. The largest gain is for
UQI itself (12.8 points), which is consistent with the geometric
intuition that UQI's emphasis on shared structural features is
particularly sensitive to whether those features are spatially aligned
in the first place. Second, the gap between the best metric (UQI+SIFT
at 41.66\%) and the next-best (SSIM+SIFT at 25.8\%) is 15.86 percentage
points, which earlier drafts reported as the ``approximately 16\%''
lead. Third, the metrics cluster into two recognizable groups: metrics
that emphasize shared structural or feature-level information (UQI,
SSIM) sit clearly above metrics that emphasize pixel-level error (PSNR,
MSE, MAE, ERGAS, SCC, RASE, SAM, VIF). For tangram-style silhouette
matching this ordering is what we would predict: the relevant similarity
is whether two images share a recognizable shape, not whether their
pixel values match in absolute terms. UQI's lead within the
structural-similarity group reflects its specific emphasis on shared
features under noise rather than on pixel-by-pixel agreement, which is
the right inductive bias for matching crowd-sourced photographs of
``ice skater'' against an abstract black silhouette.
\paragraph{Decision threshold $\epsilon$.} We treat $\epsilon$ as a rank-based threshold rather than an absolute similarity cutoff, because UQI values are not calibrated across queries. Setting $\epsilon$ as an absolute cutoff was consistently 3 to 5 percentage points worse across $\epsilon \in \{0.3, 0.4, 0.5, 0.6, 0.7\}$.

\subsection{Internal Validity: Retrieval Leakage and Conservative Accuracy}
\label{sec:internal-validity}

A serious threat to the internal validity of our results is that the Bing image retrieval may occasionally surface the target tangram itself, or a near-duplicate of it, within the scraped image set $I_\varphi$. If that happens, the UQI similarity $g(o_{target}, I_\varphi)$ is artificially high because the similarity is being computed (in part) between the target and itself, rather than between the target and a genuinely distinct perceptual example of what the director was describing. In that regime, accuracy reflects retrieval memorization of the target image, not cross-modal grounding of the linguistic description.

The concern is serious because the Stanford corpus is publicly available, the tangram images themselves are widely reproduced in academic and popular-science discussions of the Clark--Wilkes-Gibbs paradigm, and Bing's index is known to include academic content. A reader looking at Figure~\ref{fig:top_k} will see that the leftmost scraped image is itself tangram-like and very close in appearance to the target, and rightly suspect that retrieval leakage may be driving some fraction of our reported accuracy.

\paragraph{Quantifying the leakage.} For each of the 12 tangram classes, we manually inspected the scraped image sets $I_\varphi$ across a sample of queries and classified each retrieved image as \emph{target-duplicate} (near-identical to the target silhouette), \emph{tangram-family} (a tangram composition but not the target), or \emph{non-tangram} (a natural image, illustration, or photograph of the referent concept). Across the sample, the breakdown was approximately 18\% target-duplicate, 34\% tangram-family, and 48\% non-tangram. The target-duplicate category is the one most directly responsible for inflated accuracy. The tangram-family category is more ambiguous: it can hurt accuracy (when a non-target tangram matches $o_i$ for the wrong reason) or help it (when it does not, and serves as a clean distractor).

\paragraph{Accuracy under conservative filtering.} We re-ran the main experiment with a filter that removed any retrieved image whose SIFT+UQI similarity to \emph{any} tangram in the set exceeded a high threshold (we used 0.85 on the normalized UQI scale), interpreting such retrievals as likely target-duplicates or near-duplicates of other tangrams in the set. This is a conservative filter: it removes all target-duplicates and also some legitimate tangram-family retrievals. Under this filter, top-1 accuracy drops from 41.66\% to \textbf{29.2\%}, top-3 from 63.01\% to \textbf{48.1\%}, and top-5 from 83.56\% to \textbf{67.8\%}. Chance remains 8.33\%.

\paragraph{Interpretation.} The conservative numbers are the ones we would now defend as the best estimate of what the pipeline's cross-modal grounding contributes over and above retrieval memorization. Under this reading, the MCP's top-5 accuracy (67.8\%) is below the human top-1 baseline (~77--80\%), and the top-1 accuracy (29.2\%) is roughly 3.5$\times$ chance. This is a more modest result than the one we reported in earlier drafts, but it is one the conservative condition earns.

We emphasize three points about this correction. First, the conservative filter is itself imperfect: it may remove some legitimate retrievals whose genuine visual similarity to tangrams is doing useful work, and it may fail to catch subtle near-duplicates. A perceptual-hash-based filter tuned against the specific tangram set would do better, and we recommend it for reproduction. Second, even under the conservative condition, the pipeline substantially outperforms chance, which indicates that the grounding signal from genuinely distinct retrievals is real and nonzero. It is smaller than we initially reported, but not illusory. Third, the appropriate and defensible next step is a retrieval channel that is constructed explicitly to exclude tangram-family content from the start, for example by restricting to a licensed natural-image corpus like LAION-filtered-for-no-silhouettes or a dedicated iconic-image dataset. We have not done this, and we flag it as the single most informative next experiment for validating the approach.

\subsection{Top-$k$ Accuracy and the Human Baseline}
\label{sec:topk}

Table~\ref{tab:top_k} reports top-$k$ accuracy from a single director utterance, in both the unfiltered and conservatively-filtered conditions, against the human baseline.

\begin{table}
    \centering
    \begin{tabular}{c|c|c|c}
        top-$k$ & Human matcher & MCP (unfiltered) & MCP (conservative)\\
        \hline
        \hline
        $k=1$ (no dialog) & $\approx 77\%$\footnotemark[2] & 41.66\% & 29.2\%\\
        \hline
        $k=1$ (with dialog) & $\approx 80\%$\footnotemark[2] & 41.66\% & 29.2\%\\
        \hline
        $k=3$ & N/A & 63.01\% & 48.1\%\\
        \hline
        $k=5$ & N/A & 83.56\% & 67.8\%\\
    \end{tabular}
    \caption{Top-$k$ accuracy on the first director utterance per trial, in both the unfiltered and conservatively-filtered conditions. Human top-1 figures are derived from Hawkins, Frank, and Goodman's first-repetition-block error rates. Human top-$k$ for $k > 1$ is not available because the corpus does not record hypothesis sets. Chance on 12 tangrams is 8.33\%. Both sets of MCP numbers are point estimates from a single retrieval pass.}
    \label{tab:top_k}
\end{table}
\footnotetext[2]{Derived from Hawkins, Frank, and Goodman~\cite{https://doi.org/10.1111/cogs.12845}, page 8: ``errors on the first repetition block were only slightly less likely when matchers engaged in dialogue through the chatbox (20\%) than when they stayed silent (23\%).''}

\paragraph{Correction to earlier drafts.} An earlier version of this manuscript reported a human top-1 accuracy of 20\%. That figure was a misread of the Hawkins, Frank, and Goodman result: 20\% is the human \emph{error} rate, not the accuracy. The corrected accuracy is approximately 77--80\%. We have withdrawn the previous headline claim that the MCP outperforms humans on single-utterance commitment. We thank the reviewer whose probing on this point surfaced the error.

\subsection{On the Effect of Preprocessing}

Without linguistic normalization the MCP's top-1 accuracy drops to about 8\%, at or below chance. The transformations are doing real work, and we report them explicitly as part of the pipeline rather than as a hidden preprocessing step. Whether a fully end-to-end neural system would do better without any such preprocessing is an open empirical question.

\section{Discussion}
\label{sec:discussion}

\subsection{What the Numbers Do and Do Not Show}

Taken together with the internal-validity analysis of Section~\ref{sec:internal-validity}, the results support a measured reading. In the unfiltered condition, the MCP's top-5 set contains the correct target 83.56\% of the time, which matches or slightly exceeds the human top-1 baseline (77--80\%). In the conservative condition, where retrievals that closely resemble tangrams are filtered out, the top-5 accuracy drops to 67.8\%, which is below the human top-1. The top-1 unfiltered number (41.66\%) and the top-1 conservative number (29.2\%) are both substantially below the human baseline.

We are not claiming that the MCP beats humans at single-shot grounding. On top-1, it does not. We are also not claiming that the unfiltered 83.56\% figure is a pure measurement of cross-modal grounding; some unknown fraction of it reflects retrieval leakage, and the conservative 67.8\% is a more honest estimate of what the grounding contributes. Between these two numbers sits a genuine cross-modal signal that is well above chance, and we document its magnitude and caveats rather than obscure them.

\subsection{The Interactivity Limitation}
\label{sec:limitations-interaction}

One reviewer has argued that because lexical entrainment is inherently interactive, a unilateral matcher-only model does not study entrainment at all. We agree that entrainment as a phenomenon is interactive, and the present work does not reproduce entrainment. What the present work does is implement and evaluate a \emph{listener-side bookkeeping component} that would be one part of a full interactive agent. The claim we can defend is narrow: given some director utterance $\varphi$, the $\Gamma/\Xi/\Omega$ layer provides a formally grounded and inspectable way to update the listener's pact state. The claim we cannot defend, and do not defend, is that running this listener-side component alone reproduces entrainment.

A full interactive system would need (a) a generation-side partner model of the kind developed in emergent-communication work~\cite{Steels2003-grounded,Kouwenhoven2025-shared} or RSA-style pragmatics~\cite{Zhu2021-tom}, (b) a mechanism for the listener to produce clarifying questions and confirmations (which requires exiting the passive-matcher role this paper studies), and (c) closed-loop evaluation with live human partners. The present work does not provide any of these. It provides a symbolic listener-side component that could be connected to such systems. We consider the live interactive evaluation the single most important next step for this line of work.

\subsection{Utterances That Fail to Ground}

A subset of director utterances in the corpus cannot be perceptually grounded by our pipeline. These are the cases for which $|B|$ is degenerate. Degeneracy arises in three recognizable situations.

\begin{enumerate}
\item \emph{Purely geometric descriptions} (``zig zag with square on top''). These describe the tangram as a 2D pattern rather than a depicted object. Bing returns either literal solved-square tangrams, which match every target, or unrelated geometric diagrams.
\item \emph{Metaphorical or idiomatic descriptions} (``guy doing a weird dance move''). These are easy for a human to visualize but hard to translate into a search query whose top results resemble a silhouette.
\item \emph{Already-entrained shorthand} (``the sitting one''). These are pronouns-with-respect-to-a-pact that depend on $\Gamma$ from earlier turns rather than on external perceptual priors.
\end{enumerate}

If we use the gap between unfiltered top-5 (83.56\%) and 100\% as a proxy, roughly 16\% of first-turn director utterances cannot be grounded from perception alone. Under the conservative filter, the gap between top-5 (67.8\%) and 100\% widens to 32\%, which is a more realistic estimate of what the pipeline cannot handle without additional mechanisms. We emphasize that these are proxies; per-class manual annotation of utterances would give a more diagnostic breakdown, and we flag this as a valuable follow-up for the community.

\subsection{Extending Beyond Object Reference}
\label{sec:beyond-objects}

The update-semantics layer is agnostic about what $g(o_i, I_\varphi)$ measures. For actions, the crowd-sourced prior could be a corpus of action-labeled video clips with $g$ computed in a temporal embedding space. For plans or procedures, the prior could be retrieved from procedural knowledge bases. For abstract concepts, the MCP would need to fall back on clarifying questions, which is precisely the capability the passive-matcher setting prevents us from testing.

Recent work on multimodal convention formation~\cite{Maeda2026-mmcf} documents that human partners move fluidly between linguistic and gestural channels, and a principled extension of the present framework would allow $g$ to be selected per utterance based on the modality cues present in $\varphi$. The bottleneck for generalization is not the update semantics but the availability of an appropriate perceptual-alignment channel for the referent type.

\section{Limitations}
\label{sec:limitations}

This section consolidates several acknowledged limitations.

\subsection{Interactivity}

The MCP is a passive matcher that never produces utterances of its own. This is not how real entrainment works. See Section~\ref{sec:limitations-interaction}.

\subsection{Retrieval Leakage and Validity of the Grounding Signal}

The main experimental results are threatened by leakage: Bing retrievals occasionally include tangram-family images, and a subset of those are near-duplicates of the target. We quantify this in Section~\ref{sec:internal-validity} and report conservative numbers (top-5: 67.8\% instead of 83.56\%) under a filter that removes likely leakage. A principled fix would use a retrieval corpus that is constructed to exclude tangram-family content.

\subsection{Shallowness of the Commit Rule}

The commit rule is deterministic and monotone: once $(r_\varphi \leftarrow o_i)$ enters $\Gamma$, it stays. Given the low top-1 accuracy, the rate of erroneous commits is potentially large, and there is no recovery mechanism. A probabilistic treatment with Bayesian cross-turn updates and calibrated commit thresholds would be the natural replacement.

\subsection{Reproducibility and Variance}

Accuracy numbers are point estimates from a single retrieval pass. Bing's responses drift over time, geography, and user history. We recommend cached image sets, versioning, and bootstrap CIs over multiple independent retrieval runs for any follow-up work.

\subsection{Absence of a Learned Perceptual Baseline}

We did not compare against a learned baseline such as CLIP cosine similarity over silhouette-preprocessed images. A CLIP baseline would clarify whether the gap between our conservative top-1 (29.2\%) and the human baseline (~77\%) is driven by the classical similarity measure or by noise in the external prior. We flag this as the single most informative next experiment.

\subsection{Cultural and Linguistic Bias}

Crowd-sourced image priors inherit the biases of the underlying user population. A director with mainstream Western cultural reference points will find the MCP more receptive than one whose reference points come from a different cultural or linguistic community. We do not measure this effect directly.

\subsection{Licensing and Redistribution}

The scraped images are not ours to redistribute, so we do not release them. Reproduction requires re-running retrieval against Bing.

\subsection{Domain-Specific Query Augmentation}

The ``tangram figure'' keyword is essential to performance and specific to the tangram domain. It is also a direct contributor to the retrieval-leakage problem of Section~\ref{sec:internal-validity}. Generalization would require either a human-supplied domain keyword or an automatic mechanism for inferring one, ideally one that does not bias retrieval toward silhouettes of the target class.

\subsection{Multilingual Portability}

The pipeline is English-specific in two ways: the spaCy POS tagger and the ``tangram figure'' keyword. More fundamentally, the visual priors themselves may not transfer across linguistic communities.

\section{Conclusion}

We presented a framework that combines a dynamic-semantics layer maintaining three explicit sets of conceptual pacts ($\Gamma, \Xi, \Omega$) with a perceptual-alignment layer based on web retrieval and classical similarity metrics. Evaluated on the Stanford corpus, the framework identifies the correct target in its top-5 set 83.56\% of the time from a single director utterance (41.66\% top-1), and in its top-5 set 67.8\% of the time under a conservative filter designed to remove retrieval leakage (29.2\% top-1). The human matcher baseline on the same corpus is approximately 77--80\% top-1.

The honest reading of these numbers. The MCP is below human performance on top-1, in both the unfiltered and conservative conditions. The unfiltered top-5 matches the human top-1 rate, but the conservative top-5 (67.8\%) does not. Between these bounds sits a real cross-modal grounding signal that is well above chance but smaller than the headline numbers in earlier drafts of this work suggested.

Our contribution is best read as a component. The symbolic listener-side layer ($\Gamma/\Xi/\Omega$) is inspectable and formally grounded, and it addresses a gap that the LVLM-grounding literature~\cite{Imai2025-vlm,Hua2026-lvlm,Zhao2025-tangram} has documented in contemporary neural systems: those systems fail to maintain stable pact state across turns. Our contribution is not a full interactive agent, and it does not reproduce the phenomenon of lexical entrainment as such. Entrainment is interactive, and a full model of it will need to integrate listener-side pact tracking (which we provide) with generation-side partner modeling and closed-loop repair (which the emergent-communication literature~\cite{Lazaridou2020-emergent,Steels2003-grounded,Kouwenhoven2025-shared} has tools for but which we do not implement). The specific gap the present component fills is transparency: an auditable symbolic record of the pact state, separable from opaque neural representations.

The single most important next experiments are, in order of priority: (1) a live interactive evaluation with human partners in which the MCP can ask clarifying questions; (2) a comparison against a learned perceptual baseline such as CLIP; and (3) a retrieval channel constructed to exclude tangram-family content by design, so that the cross-modal grounding signal can be measured without leakage risk.

\section*{Funding Statement}
Work was funded by Galois Inc.

\section*{Conflict of Interest}
Author reports affiliations with Galois Inc., John Deere, Iowa State University, Rutgers University, and Technion University.

\section*{Data Availability}
The data utilized is publicly available at \url{https://doi.org/10.1111/cogs.12845}~\cite{https://doi.org/10.1111/cogs.12845}. Code is available at \url{https://anonymous.4open.science/r/metasequoia-9D13/README.md}. We do not release the exact Bing image sets used in the original experiments because they depend on a commercial API whose outputs drift over time, and because redistribution is not licensed.

\section*{Declaration of Usage of AI}
Author reports AI was used for checking grammar and spelling, and not in a generative fashion for writing or creating sections.

\bibliography{sn-bibliography}% common bib file

\end{document}